\documentclass[prd,preprint,showpacs,groupedaddress]{revtex4-1}
\usepackage{amsmath}
\usepackage{amsfonts}
\usepackage{amssymb}
\usepackage{geometry}
\usepackage{natbib}
\usepackage[english]{babel}
\usepackage{graphicx}
\usepackage{epstopdf}
\usepackage{subfigure}
\usepackage{caption}
\usepackage{multirow}
\usepackage{indentfirst}
\usepackage{mathrsfs}
\usepackage[colorlinks,linkcolor=blue,anchorcolor=blue,citecolor=blue]{hyperref}
\begin{document}

  \setlength{\parindent}{2em}
  \title{ The cosmological constant from space-time discreteness}
  \author{Qi-Qi Fan}\author{Cong Li}\author{Hao-Ran Zhang}\author{Peng-Zhang He}
  \author{Jian-Bo Deng}\thanks{Jian-Bo Deng: dengjb@lzu.edu.cn}
  \affiliation{Institute of Theoretical Physics $\&$ Research Center of Gravitation, Lanzhou University, Lanzhou 730000, China}
  \date{\today}

  \begin{abstract}
  We regard the background of space-time as a physical system composed of discrete volume elements at the Planck scale and get the internal energy of space-time by Debye model. A temperature-dependent minimum energy limit of the particles is proposed from the thermal motion part of the internal energy. As decreases of the temperature caused by the expansion of the universe, more and more particles would be "released" because of the change of the energy limit, we regard these new particles as a source of dark energy. The minimum energy limit also leads to a corrected number of particles in universe and a modified conservation equation. According to the modified conservation equation, an effective cosmological constant consistent with its observed value is obtained. 

  \end{abstract}


  \maketitle
  \section{Introduction}\label{sec1}
  The astrophysical data obtained from high redshift surveys of supernovae~\cite{1,2} reveals the fact that the expansion of the universe is accelerating. This acceleration is attributed to an unknown entity called dark energy (DE). In the past few decades, various modifications of general relativity have been proposed to explain the origin of it. Some of the most studied models are: the cosmological constant model, $f(R)$ gravity~\cite{3,4,5,6}, scalar-tensor theories~\cite{7,8}, quintessence models~\cite{9,10} and phantom models~\cite{11,12}, etc. Unfortunately, as one of the most successful interpretations of DE, the value of the cosmological constant $ \Lambda $ calculated by the cosmological constant model is 60-120 orders of magnitude smaller than that given by quantum field theory~\cite{13}. This indicates that there is an irreconcilable contradiction between general relativity (GR) and quantum field theory (QFT).
\par
  In order to eliminate the contradiction between these two theories, various quantum gravity theories~\cite{14,15,16,17,18,19,20} have been proposed in past few years. As an approach to find a suitable quantum gravitational theory, some physicists have considered the effect of the discreteness of space-time at a very small scale. In these theories, the fundamental description of space-time is only an approximation to manifold. For instance, causal set theory~\cite{21,22,23,24} described the space-time as a set of discrete points and developed the quantum dynamics of particles at these points. The points fixed on the background manifold are treated as the position of particles, therefore, the discreteness of the points will affect the trajectory of the particles at a tiny scale. A recent work~\cite{25} completed by Alejandro Perez and Daniel Sudarsky interpreted the cosmological as the accumulation of small violating the energy conservation which derived from the discreteness at the Planck scale. In these studies, the background of the universe was considered to have discrete microstructures, we follow this view and further hypothesize that the discrete points on the background have properties similar to particles, thus one can study the dynamics or thermodynamics of the background. This assumption is equivalent to figure out the type of interaction between volume elements (voxels) of space-time, not just whether it is discrete or continuous. If there is a suitable micromodel to describe the background, it will provide possibilities for researching the energy exchange between background and the particles on the background, this is exactly the work studied in this article. 
\par  
  In this paper, we regard the background as a thermodynamic system with strong interaction and discrete structure at the Planck scale, so the Debye model~\cite{26} can be used. By calculating the internal energy expression of the background under thermal equilibrium conditions, we obtain a temperature-dependent energy limit, considering the influence of the thermal motion of the background on the particles, this energy limit is assumed as the lowest energy of the particles. As the energy limit decreases, more and more particles contribute to gravity. As time goes on, this new energy will dominate the expansion of the universe. We regard it as a source of dark energy and calculate the cosmological constant corresponding to energy density of these new particles, which is consistent with the observation value of the cosmological constant~\cite{27}.
  
  The paper is organized as follows. Sec.~\ref{sec2} briefly review the method of obtaining the Debye model by canonical ensemble theory, then replace the ordinary atoms in solid with Planck-sized voxels and obtain the expression of the internal energy in the background of space-time. In Sec.~\ref{sec3}, the relationship between the number of the "real" particles and the total number of particles in universe is obtained using the lowest energy limit. Then we get a new conservation equation and estimated the value of the cosmological constant in Sec.~\ref{sec4}. Finally Sec.~\ref{sec5} gives some conclusions and discussions. We will use natural units in which $\hbar=c=k_{B}=1$ in this article.

\par

\section{ the Debye model of the background of the universe}\label{sec2}
 Physically, to take background of space-time as an object of research, we need an appropriate model to describe this system. Assuming space-time is discrete, meaning that space-time has internal structure, so we can study it with the theory of statistical physics. Intuitively, a system used to describe the background of the universe must not be a thermodynamic system with weak interactions like ideal gas, because such a system is not stable enough, the system will "tear" when the energy is too high. Even a rest particle may also move chaotically while be embedded in such a background. Therefore, the background system should be dense, with strong interaction between discrete voxels. Fortunately, in statistical physics, there is an appropriate example to describe such a system, that is the Debye model.
\par
 In general solids, there is strong interaction between the atoms,  which makes the atoms have their own equilibrium positions and do micro-vibration near their equilibrium positions. Assuming that there are $N$ atoms in the solid and each atom has 3 degrees of freedom, then the solid has 3$N$ degrees of freedom. The total potential energy $ \phi $  can be simplified as
  \begin{equation}\label{eq: potential energy}
  \begin{aligned}
   \phi=\phi_{0}+\sum_{l}\left(\dfrac{\partial\phi}{\partial\xi_{l}}\right)\xi_{l}+\frac{1}{2}\sum_{l,s}\left(\dfrac{\partial^{2}\phi}{\partial\xi_{l}\partial\xi_{s}}\right)_{0}\xi_{l}\xi_{s},
  \end{aligned}
  \end{equation}
  where $ \xi_{l} $ is the displacement of the l-th degree of freedom from the equilibrium, and $ \phi_{0} $ is the potential energy of the system when all atoms are in equilibrium. In addition to potential energy, each degree of freedom has corresponding microvibration energy. Therefore, the total energy of the system can be obtained
  \begin{equation}
  \begin{aligned}
    \label{eq:The total energy}
    E=\sum_{l=1}^{3N}\dfrac{p^{2}_{\xi_{l}}}{2m}+\frac{1}{2}\sum_{l,s}\left(\dfrac{\partial^{2}\phi}{\partial\xi_{l}\partial\xi_{s}}\right)_{0}\xi_{l}\xi_{s}+\phi_{0},
  \end{aligned}
  \end{equation}
  where $ p^{2}_{\xi_{l}}/2m $ is the kinetic energy of l-th degree of freedom. In canonical coordinates, the above formula can be expressed as the sum of 3$N$ independent simple harmonic motions. As we all know, in quantum mechanics, the energy of such 3$N$ independent simple harmonic motion can be expressed as
   \begin{equation}
   \begin{aligned}
    \label{eq:3}
   E=\phi_{0}+\sum_{i=1}^{3N}\hbar\omega_{i}\left(n_{i}+\dfrac{1}{2}\right),n_{i}=0,1,2,\cdots,
   \end{aligned}
   \end{equation}
  where, $ \omega_{i} $ is the normal frequency and $ n_{i} $ is quantum number describing the i-th simple harmonic motion. In this way, we get a form of energy that we can work on.
  
  In statistical physics, it is convenient to use the canonical ensemble theory to study systems with strong interactions. The partition function of the above system can be expressed as
  \begin{equation}
   \begin{aligned}
    \label{eq:4}
    Z=e^{-\beta\phi_{0}}\prod_{i}\frac{e^{-\frac{\beta\hbar\omega_{i}}{2}}}{1-e^{-\beta\hbar\omega_{i}}}.
   \end{aligned}
   \end{equation}
 According to the canonical ensemble theory, the internal energy of the system can be obtained
  \begin{equation}
  \label{eq:5}
   U=-\dfrac{\partial}{\partial\beta}\log Z=U_{0}+\sum_{i=1}^{3N}\frac{\hbar\omega_{i}}{e^{-\beta\hbar\omega_{i}}-1},
  \end{equation}
  where 
  \begin{equation}\label{6}
   U_{0}=\phi_{0}+\sum_{i}^{3N}\frac{\hbar\omega_{i}}{2}.
  \end{equation}
  Generally, $ \phi_{0} $ is negative, and its absolute value is greater than zero energy, so we get a negative binding energy $ U_{0} $ which is independent of temperature. The second term on the right side of~\eqref{eq:5} represents the energy of the thermal motion of atoms. To get the result  of internal energy specifically, we need to find the frequency spectrum of the simple harmonic motion. Debye regarded the solid as a continuous elastic medium and gave a spectrum of the system. The number of simple harmonic motions in the range of $ \omega $ to $ \omega+d\omega $ is
  \begin{equation}
  \begin{aligned}
  \label{eq:7}
  D(\omega)d\omega=\frac{V}{2\pi^2}\left(\dfrac{1}{c_{1}^3}+\dfrac{2}{c_{2}^3}\right)\omega^2\rm d\omega,
  \end{aligned}
  \end{equation}
  where $ c_{1} $ and $ c_{2} $ represent the propagation velocity of longitudinal wave and transverse wave respectively. In addition, since such a system has only 3$N$ simple harmonic vibrations, there must be an upper frequency limit, so the following formula holds
   \begin{equation}
  \begin{aligned}\label{eq:8}
  \int_{0}^{\omega_{D}}B\omega^2\rm d\omega=3N,
  \end{aligned}
  \end{equation}
  for this, we get
  \begin{equation}
  \begin{aligned}\label{eq:9}
  \omega_{D}^3=\dfrac{9N}{B},
  \end{aligned}
  \end{equation}
  where
  \begin{equation}
  \begin{aligned}\label{eq:10}
  B=\frac{V}{2\pi^2}\left(\dfrac{1}{c_{1}^3}+\dfrac{2}{c_{2}^3}\right).
  \end{aligned}
  \end{equation}
  As can be seen from~\eqref{eq:9}, $ \omega_{D} $ is related to the density of atoms and the velocity of the elastic wave, this fact is important to understand the relationship between scale and $\omega_{D}$  when we apply~\eqref{eq:9} to the background of univese. Using the Debye spectrum,~\eqref{eq:5} can be expressed as
   \begin{equation}\label{eq:11}
   U=U_{0}+B\int_{0}^{\omega_{D}}\frac{\hbar\omega^3}{e^{\frac{\hbar\omega}{kT}}-1}\rm d\omega.
   \end{equation} 
  In statistical physics, this integral has different results under different temperature. When the temperature of the system is much higher than the characteristic temperature $\theta_{D}=\omega_{D}$, i.e. $ T\gg\theta_{D}$, the result of~\eqref{eq:11} is
  \begin{equation}\label{eq:12}
  U=U_{0}+3NT.
  \end{equation}
 At low temperatures $T\ll\theta_{D}$, the internal energy of the system is approximately equal to
 \begin{equation}\label{eq:13}
 U=U_{0}+3N\dfrac{\pi^4}{5}\dfrac{T^4}{\theta_{D}^3}.
 \end{equation}
 the result in \eqref{eq:13} indicates that the thermal motion part of internal energy is limited by the characteristic temperature $ \theta_{D} $ at low temperature, this property guarantees the stability of strongly interacting systems. Similarly, stability can also be guaranteed when we apply the above formula to the background of the universe.
  
 Now, we have reviewed the process to get the internal energy expression of the strongly interacting system in statistical physic. As mentioned above, we regard space-time as a physical object similar to a strongly interacting system, therefore,the expression in ~\eqref{eq:11} will be used to represent the internal energy of the background. We will see that the thermal motion part of internal energy plays a crucial role in the production of dark energy.   
 \section{Minimum energy limit}\label{sec3}

 Let us study~\eqref{eq:11} from the view of cosmology. As mentioned above, the space-time is tread as a discrete system which has strong interaction between each small voxel like general solids, thus the internal energy expression in~\eqref{eq:11} can also be used to represent the internal energy of space-time. The first term on the right side of~\eqref{eq:11} $ U_{0} $ is the binding energy of the background of the universe. We assume that, in general, this part of the energy does not interact with the particles moving on the background, so it does not contribute to gravity. The last term of~\eqref{eq:11} represents the energy of the thermal motion of the voxels, $T$ is the temperature when the background and particles are in thermal equilibrium. This means that there is energy exchange between the background and the particles. Assuming space-time is composed of $ N $ voxels, then the thermal motion energy of one voxel is 
  \begin{equation}\label{eq:14}
  \varepsilon=\dfrac{3}{5}\pi^4\dfrac{T^4}{\theta_{D}^3},
  \end{equation}
  where $\theta_{D}$ is determined by the number density of voxels and the parameter $B$ in~\eqref{eq:9}. In this article, the scale of voxels is at the order of the Planck scale $\ell_{p}$, thus the corresponding number density $n=T_{p}^3$, where $T_{p}$ is the Planck tempetature, then~\eqref{eq:9} can be re-expressed as
  \begin{equation}\label{eq:15}
  \theta_{D}^3=\dfrac{18\pi^2}{\dfrac{1}{c_{1}^2}+\dfrac{2}{c_{2}^2}}T_{p}^3,
  \end{equation}
  we suppose that the speed of transverse wave $c_{1}$ and longitudinal waves $c_{2}$ is equal to the speed of light,i.e. $c_{1}=c_{2}=c=1$. In fact, as long as $c_{1}$ and $c_{2}$ is in the order of $c$, the choice of the value of them has little effect on the result. As we can see from the above formula, $\theta_{D}\approx T_{p}$, this means that the relationship $T\ll\theta_{D}$ is always hold during most period of the universe, so~\eqref{eq:13} is  the correct expression for the background of the universe. Substituting the result of~\eqref{eq:15} into~\eqref{eq:14}, a more specific energy expression can be obtained
  \begin{equation}\label{eq:16}
  \varepsilon=\dfrac{\pi^2}{10}\dfrac{T^4}{T_{p}^3},
  \end{equation}
  which means that, on the cosmological scale, there is a vibration  is in the order of $T^4/T_{p}^{3}$ in every point around the background. Different from the general case where only material exists, the vibration of voxels in background will have an important effect on the movement of particles because the particles are embedded in the space-time. Therefore, it is reasonable to believe that there is a minimum value of the energy of the particles given by the vibration of the background. The particles may not show up below this energy limit.
   
\par
  In order to reflect the energy exchange caused by the minimum energy limit, it is useful to calculate the number of particles. The period is paid attention to during which the reheating of the electroweak transition is completed, the temperature of the  universe is much higher than that of the static mass of most particles, where radiation is dominant. According to statistical physics, the state density of relativistic particles is given by
  \begin{equation}\label{eq:17}
  D(\varepsilon)d\varepsilon=\dfrac{V}{2\pi^2}\varepsilon^{2}\rm d\varepsilon,
   \end{equation}
  assuming that the particles follow the Boltzman distribution, in the presence of a minimum energy limit, the new number of particles is expressed as 
  \begin{equation}\label{eq:18}
  \widetilde{N}=N-\int_{0}^{\gamma\varepsilon_{c}}D\left(\varepsilon\right)e^{-\alpha-\beta\varepsilon}\rm d\varepsilon,
 \end{equation}
  where $\gamma$ is coupling constant that indicates the correlation of particles and space-time, and the value of $\gamma$ is around 1. $\varepsilon_{c}$ is the characteristic energy obtained by~\eqref{eq:17}. To get the result of the above formula, the expression of $e^{-\alpha}$ must be known. It can be gotten from
  \begin{equation}
  \begin{aligned}\label{eq:19}
  N=\int_{0}^{\infty}D\left(\varepsilon\right)e^{-\alpha-\beta\varepsilon}\rm d\varepsilon.
  \end{aligned}
  \end{equation}
  Substituting the~\eqref{eq:17} into~\eqref{eq:19}, one obtains  
  \begin{equation}\label{eq:20}
   e^{-\alpha}=\pi^2\dfrac{N}{V}\dfrac{1}{T^3}.
  \end{equation}
  Then one gets the number of particles with energy above $\gamma\varepsilon_{c}$ from~\eqref{eq:18} is
  \begin{equation}\label{eq:21}
  \widetilde{N}=N-N\left[e^{-\frac{\widetilde{\gamma} T^3}{T_{p}^{3}}}\left(-\dfrac{\gamma T^6}{2T_{p}^{6}}-\dfrac{\gamma T^3}{T_{p}^{3}}-1\right)+1\right],
  \end{equation}
  where $\widetilde{\gamma}=\frac{\pi^2}{10}\gamma$. Planck temperature is the upper limit of temperature allowed, and even at the beginning of the time we studied, the temperature $T$, which approximates 100 GeV, is much lower than the Planck temperature $T_{P}\approx10^{19}$ GeV. Ignoring the small terms in~\eqref{eq:21}, a simplified modified number of particles is obtained
  \begin{equation}
  \begin{aligned}\label{eq:22}
  \widetilde{N}\approx Ne^{-\frac{\widetilde{\gamma} T^3}{T_{p}^{3}}}.
  \end{aligned}
  \end{equation}
  
  The above formula means that the "real" particles in the universe are not the number of all particles (if there is a certain number of particles at the beginning of the Big Bang and it is roughly unchanged at a latter time), but partly cut off by a very small energy limit which changes with $T^{3}$. As the temperature decreases, more and more particles are released, this process is accompanied by the accumulation of energy. When $T\rightarrow0$ and $\widetilde{N}\rightarrow N$, this energy accumulation process is over. In our study, this part of the energy is considered as the source of dark energy, and according to our calculations, we will see that the energy density of this part of energy is consistent with the energy density represented by cosmological constant observed.
  \section{Dark Energy}\label{sec4}
  
  At the end of the previous section, we obtained a corrected the number of particles. In this case, the corresponding particle number density is
  \begin{equation}\label{eq:23}
  \widetilde{n} \propto T^{3}e^{-\frac{\widetilde{\gamma} T^{3}}{T_{p}^{3}}}.
  \end{equation}
  In the standard cosmological model, the volume of universe is proportional to $a\left(t\right)^{3}$, which means that the particle number density $n\propto a\left(t\right)^{-3}$. Therefore, a corrected Hubble parameter is obtained
  \begin{equation}\label{eq:24}
  \widetilde{H}=\frac{\dot{\widetilde{a}}}{\widetilde{a}}=-\frac{1}{3}\frac{\dot{\widetilde{n}}}{\widetilde{n}}=-\frac{\dot{T}}{T}\left(1-\gamma\frac{T^3}{T_{p}^{3}}\right)=H+\gamma\frac{T^2\dot{T}}{T_{p}^{3}},
  \end{equation}
  where $H$ is the Hubble parameter in standard cosmological model. Now, we pay attention to the influence of such a Hubble parameter on the conservation of energy momentum tensor. Under the flat Friedmann-Lemaître-Robertson-Walker metric
  \begin{equation}\label{eq:25}
  ds^{2}=-dt^{2}+a\left(t\right)^{2}d\vec{x}^{2},
  \end{equation}
  the conservation equation $\triangledown^{\mu}T_{\mu\nu}=0$ takes the form
  \begin{equation}
  \begin{aligned}\label{eq:26}
  \dot{\rho}+3H\left(1+\omega\right)\rho=0,
  \end{aligned}
  \end{equation}
  where, $T_{\mu\nu}$ is the energy momentum tensor of the perfect fluid, $\rho$ refers to energy density (including radiation and matter) and $\omega$ is the Equation-of-State (EoS) parameter of the perfect fluid. Replacing the Hubble parameters $H$ with~\eqref{eq:24}, one gets a corrected conservation equation as
 \begin{equation}
  \begin{aligned}\label{eq:27}
  \dot{\rho}+3H\left(1+\omega\right)\rho=-\widetilde{\gamma}\left(1+\omega\right)\frac{T^2}{T_{p}^{3}}\dot{T}\rho.
  \end{aligned}
  \end{equation}
  Comparing with~\eqref{eq:26}, a term that represents the energy flow density appears on the right side of the new conservation equation, defining 
  \begin{equation}\label{eq:28}
   \dot{\rho}_{\Lambda}=-\widetilde{\gamma}\left(1+\omega\right)\frac{T^2}{T_{p}^{3}}\dot{T}\rho,
  \end{equation}
  on the large scale of the universe, the decrease of temperature leads to a negative $\dot{T}$, if there is always $-1<\omega<\frac{1}{3}$ in every period of the universe, $\dot{\rho}_{\Lambda}$ is always positive or close to zero. This means that this part of energy will gradually accumulate in some way and contribute to gravity, which we tend to treat it as the source of dark energy.
     
  Now, let us caculate the energy density accumulated by $\dot{\rho}_{\Lambda}$, then we can get the following integral
  \begin{equation}\label{eq:29}
  \rho_{\Lambda}=-\int_{t_{reh}}^{t_{0}}\widetilde{\gamma}\left(1+\omega\right)\frac{T^2}{T_{p}^{3}}\dot{T}\rho \rm dt=-\int_{T_{reh}}^{T_{0}}\widetilde{\gamma}\left(1+\omega\right)\frac{T^2}{T_{p}^{3}}\rho \rm dT
  \end{equation}
 where, $t_{reh}$ is the end time of the reheating period after the inflation caused by the electroweak transition, and $t_{0}$ is the time of the current universe. $T_{reh}$ and $T_{0}$ refer to the thermal equilibrium temperature of the universe at the corresponding time respectively. At the time $t_{reh}$, the universe has just finished reheating, in which the rest mass of most particles is below the temperature $T_{reh}$ and the universe is dominated by radiation, the energy density is 
\begin{equation}\label{eq:30}
\rho = \frac{\pi^2}{30}g_{\ast}T^{4}.
\end{equation}
where $g_{\ast}\approx100$ is the effective degeneracy factor during the radiation-dominated period~\cite{28}.

At the time $t_{0}$, the current universe is dominated by the dark energy. Strictly, \eqref{eq:29} is not applicable to the non-radiation-dominated universe, bacause~\eqref{eq:22} or~\eqref{eq:23} is obtained from the condition of relativistic particles. To get the correct number of particles in the non-radiation period, one just need to replace the state density in~\eqref{eq:18}. For example, for non-relativistic particles, the state density is 
\begin{equation}\label{eq:31}
 D\left(\varepsilon\right)d\varepsilon=\dfrac{V}{4\pi^2}\left(2m\right)^{3/2}\varepsilon^{1/2}d\varepsilon.
\end{equation}
In this way, we will get a result slightly different from~\eqref{eq:29} and this result affected by temperature will be less than the period dominated by radiation. Today, since the Equation-of-State (EoS) parameter $\omega\rightarrow-1$ and the energy density is much smaller than the energy density of the early universe, therefore, $\rho_{\Lambda}\left(T_{0}\right)$ in~\eqref{eq:29} can be ignored (although it is inaccurate for today’s universe). Thus the result of $\rho_{\Lambda}$ is actually determined by the temperature at the initial moment. Substituting~\eqref{eq:30} and the value of each constant into~\eqref{eq:29}, one gets
\begin{equation}\label{eq:32}
\rho_{\Lambda}\approx\frac{\pi^4}{300}g_{\ast}\gamma\left(1+\omega\right)\frac{T_{reh}^7}{T_{p}^{3}},
\end{equation}
therefore the cosmological constant $\Lambda$ is
\begin{equation}\label{eq:33}
\Lambda=8\pi G\rho_{\Lambda}=\frac{2\pi^5}{75}g_{\ast}\gamma\left(1+\omega\right)\frac{T_{reh}^7}{T_{p}^{5}}.
\end{equation}
  This result is consistent with Perez's conclusion in~\cite{25}, but from the completely different view. A slight difference to~\cite{25} is that the initial temperature we choose is the temperature after reheating, not the temperature $T_{EW}$ of the electroweak transition. As mentioned above, $T_{reh}$ is in the order of 100 GeV, but a more accurate value of $T_{reh}$ is important for calculating the energy density of dark energy. Since $T_{reh}$ is the temperature of the universe after reheating completed, its value should be slightly lower than the temperature $T_{EW}$ because of the expansion of the universe during reheating~\cite{29,30}.  According to our calculation, if the value of $T_{reh}$ is between 20 GeV and 30 GeV, we can get a result of $\Lambda$ close to the observed value of the cosmological constant $\Lambda_{obs}$.
\section{Conclusions and discussions}\label{sec5}
  In this work, we studied the background of the universe as a physical system and ulteriorly thought that the background has the microstructure at the Planck scale. Under the assumption that the background is composed of strongly interacting voxels, we used the Debye model and the canonical ensemble theory to describe the thermal motion of these voxels. With the consideration of the influence of background on particle motion, we obtained a minimum energy limit of particles from the expression of internal energy. Just as the beach appears with the ebb tide of the sea, the particles with lower energy will gradually participate in the contribution to gravity as the energy limit drops. These newly emerged particles provide a new source of energy for the universe, which is referred to the dark energy. According to the relationship between the particle number density and the scale factor, we obtained a new conservation equation and calculated the energy density of the dark energy. The corresponding cosmological constant was also calculated which is consistent with the observation value of the cosmological constant. 
  
  The discussion in this article shows that the energy density of dark energy is largely determined by the temperature at the end of the reheating period, which means that the dark energy may mainly comes from the accumulation of the early universe. As the decreases of the energy density of other components, dark energy would gradually dominate the expansion of the universe. Furthermore, such a result comes from a minimum energy limit caused by the microstructure of space-time, which reveals that space-time may have a discrete structure at the Planck scale.

\section*{Conflicts of Interest}
  The authors declare that there are no conflicts of interest regarding the publication of this paper.

\section*{Acknowledgments}
  We would like to thank the National Natural Science Foundation of China (Grant No.11571342) for supporting us on this work.

%\section*{References}

\end{document}